\documentclass[letterpaper,titlepage,11pt]{article}
\usepackage{hyperref}
\usepackage{amssymb,amsmath,amsfonts,enumerate}
\usepackage{graphicx}
\usepackage{boxedminipage}

\usepackage{epstopdf}
\def\clock{{\count0=\time
           \divide\count0 60
           \ifnum\count0<10 0\fi\the\count0
           \multiply\count0 -60 \advance\count0 \time
           :\ifnum\count0<10 0\fi \the\count0
         }}
\newcommand{\timestamp}{{\small\vbox{\hbox{\tt\jobname.tex}
\hbox{\the\day/\the\month/\the\year, \clock}}}}

\newcommand{\beq}{\begin{equation}}
\newcommand{\eeq}{\end{equation}}
\newcommand{\ben}{\begin{displaymath}}
\newcommand{\een}{\end{displaymath}}
\newcommand{\beqa}{\begin{eqnarray}}
\newcommand{\eeqa}{\end{eqnarray}}
\newcommand{\bea}{\begin{eqnarray}}
\newcommand{\eea}{\end{eqnarray}}
\newcommand{\bean}{\begin{eqnarray*}}
\newcommand{\eean}{\end{eqnarray*}}
\newcommand{\ba}{\begin{array}}
\newcommand{\ea}{\end{array}}
\newcommand{\bi}{\begin{itemize}}
\newcommand{\ei}{\end{itemize}}

\numberwithin{equation}{section}

\begin{document}

\begin{titlepage}
\begin{flushright}
\end{flushright}
\vskip 2.cm
\begin{center}
{\bf\LARGE{Hawking Radiation from Small Black Holes}}
\vskip 0.12cm
{\bf\LARGE{at Strong Coupling and Large $N$}} 
\vskip 1.5cm
{\bf Nidal Haddad
}
\vskip 0.5cm
\medskip
\textit{Departament de F{\'\i}sica Fonamental and}\\
\textit{Institut de
Ci\`encies del Cosmos, Universitat de
Barcelona, }\\
\textit{Mart\'{\i} i Franqu\`es 1, E-08028 Barcelona, Spain}\\

\vskip .2 in
\texttt{ nidal@ffn.ub.es}

\end{center}

\vskip 0.3in

\baselineskip 16pt
\date{}

\begin{center} {\bf Abstract} \end{center} 

\vskip 0.2cm 

\noindent In a previous work an approximate static metric was found of a test black string that stretches from the boundary to the horizon of the planar Schwarzschild-AdS$_{5}$ geometry. This is the gravity dual of the Unruh state for $\mathcal{N}=4$, $SU(N)$ super Yang-Mills theory on a $4$-dimensional Schwarzschild background, at large $N$ and large 'tHooft coupling. We compute the holographic stress tensor of the gravitational solution and it turns out to possess many essential features of the Unruh state for weakly-coupled Hawking radiation, such as the appearance of a negative energy density near the black hole horizon and a positive energy density at infinity. It also confirms recent results that at leading order in $N$, the expectation value of the stress tensor in the Unruh state is finite on both the future and past horizons, and that at this order there are no flux terms as is expected in the black droplet phase.   

\end{titlepage} \vfill\eject

\setcounter{equation}{0}

\pagestyle{empty}
\small
\normalsize
\pagestyle{plain}
\setcounter{page}{1}

\newpage

\section{Introduction}
One of the interesting areas the AdS/CFT correspondence \cite{Maldacena:1998a} could explore is the area of quantum black holes, or Hawking radiation \cite{Hawking:1975a,Birrell:1982a,Jacobson:2003vx}. To do so one must look for black hole solutions in AdS spacetimes with the boundary condition that the induced metric on the AdS boundary is that of a black hole \cite{Hubeny:2009ru}. Two types of such solutions were conjectured to exist \cite{Hubeny:2009ru,Marlof:2010b}; black funnels or black droplets. Black funnels are black holes with connected horizons that extend from the boundary to horizon of the planar Schwarzschild-AdS geometry - they connect with the planar black hole horizon in a shoulder-like configuration. Black droplets, on the other hand, are black holes with disconnected horizons, that is, they extend from the boundary of AdS down to some point in the bulk where they close off (or cap off) in a smooth way before they reach the planar black hole horizon; the planar black hole gains some deformation as a result of the droplet suspended above it. Black funnels and droplets are the gravitational duals of different vacuum states of $\mathcal{N}=4$, $SU(N)$ super Yang-Mills theory on black hole backgrounds, at large $N$ and large 't Hooft coupling.
There are some physical differences between them though. Black funnels (as the horizon is connected) are dual to a deconfined plasma which is strongly coupled to the boundary black hole, that is, energy trasfer is quick between them, of order $O(N^2)$. Black droplets on the other hand (as the two horizons in the bulk are disconnected) are dual to a deconfined plasma which is coupled weakly to the boundary black hole, that is, energy transfer is slow between them, of order $O(1)$, see \cite{Hubeny:2009ru,Fischetti:2012ps} for further details. A sharp phase transition is expected between the two phases which will be mediated by critical geometries of the kind proposed in \cite{Haddad:2012ss,Emparan:2011ve,Kol:2003a}.

In general, the temperature of the boundary deconfined plasma can be different from the temperature of the boundary black hole, depending on the sizes of the planar black hole and the boundary black hole, respectively. In \cite{Hubeny:2009ru,Santos:2012he} the two temperatures were taken to be equal and so this was dual to the Hartle-Hawking vacuum state, describing thermal equilibrium betweem the plasma and the boundary black hole. In \cite{Figueras:2011va}, the authors constructed, numerically, a black droplet solution where the two temperatures are different; a boundary black hole at a finite temperature and a plasma at a zero temeperature, corresponding to a black droplet suspended above the extremal Poicare horizon of AdS. This was argued to be the dual of the Unruh or the Boulware vacuum states. In this paper we are going to work in $5$ spacetime bulk dimensions and we are going to focus on the droplet phase, on the case where the two temperatures are different. We are going to take a finite (non-zero) temperature planar black hole in the bulk and a high temperature boundary black hole. In other words, the temperature of the boundary black hole is going to be much higher than the finite temperature of the surrounding plasma. This will correspond to the Unruh vacuum state, which is the steady-state in which the black hole only radiates and does not absorb positive energy - a process of black hole evaporation. In the bulk, this correspondes to a thin and long black droplet (or equivalently a test black string) extending from the boundary to the horizon of the planar Schwarzschild-AdS$_{5}$ geometry \cite{Haddad:2012ss} (see Fig.\ref{fig:BS-BB}). As was shown in \cite{Haddad:2012ss}, in the Lorentzian section the black string (or the thin and long droplet) caps off smoothly just at the planar horizon, while it has a cone structure there in the Euclidean section, reflecting the fact that the two disconnected objects are at different temperatures.

In this work we take the bulk solution found in \cite{Haddad:2012ss} and compute from it the holographic stress tensor using the familiar prescription of \cite{Balasubramanian:1999re}. By the AdS/CFT correspondence this classically computed stress tensor at the AdS$_{5}$ boundary is equivalent to the expectation value of the renormalized stress tensor, in the Unruh state, of $\mathcal{N}=4$, $SU(N)$ super Yang-Mills theory on a fixed $4-$dim Shcwarzschild black hole, at large $N$ and large 't Hooft coupling. We find that this energy-momentum tensor shares many features with stress tensors computed for weakly coupled Hawking radiation \cite{Birrell:1982a}. For example, it has the essential feature of Hawking radiation that there are negative energy densities near the black hole horizon and positive ones at infinity \cite{Birrell:1982a,Davies:1975b,Page:1982a}. This is the manifestation of particle creation, one particle with negative energy enters the black hole and its partner with positive energy escapes to infinity. This is the way a black hole loses mass, and evaporates - by absorbing negative energies. We find also that it is covariantly conserved and that it satisfies the correct trace anomaly \cite{Balasubramanian:1999re,Christensen:1977a,Henningson:1998gx}. Yet as our boundary theory is strongly coupled we of course should expect also some differences from the weakly-coupled cases studied extensively in the past. In this regard we find, in agreement with \cite{Figueras:2011va,Figueras:2013jja,Fischetti:2013hja}, that at leading order in $N$, the stress tensor is finite everywhere (in the Unruh state). In particular, we find it finite on both the future and past horizons. We also find that at this leading order in $N$ the stress tensor has no flux terms, that is, there is no radiation from the boundary black hole to infinity. The latter point confirms that black droplets are, indeed, dual to a plasma which is weakly coupled to the black hole, in the sense that a flux term, or a radiation term from the black hole, would have made faster the process of exchange of heat between the black hole and the external plasma.

The paper is organized as follows. We begin in section \ref{sec:1} by introducing the bulk metric, found in \cite{Haddad:2012ss}, for a static test black string which extends from the boundary to the horizon of the planar black hole. In section \ref{sec:2} we compute the holographic stress tensor from the bulk metric. In section \ref{sec:3} we discuss the properties of the stress tensor and we compare it to stress tensors in the literature for weakly coupled Hawking radiation, and also to stress tensors for strongly coupled cases found recently. In section \ref{sec:comments} we conclude by some comments.
 
\section{Bulk metric}
\label{sec:1}
\begin{figure}[ht]
\begin{center}
\includegraphics[width=0.6\textwidth]{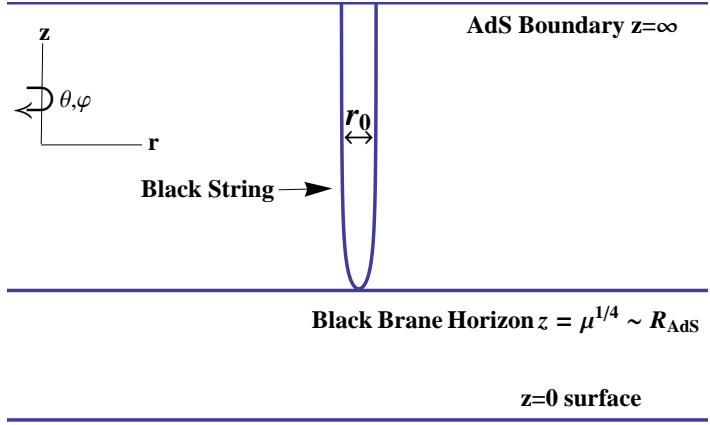}
\end{center}
\caption{This is a dipiction of the black string-black brane system. Here we have suppressed the angular directions (on $z=\text{constant}$ surfaces) $\theta$ and $\phi$, and we have kept only the radial coordinate $r$ (on those surfaces) and the holographic coordinate $z$. Here, the $r=0$ line halves the black string.
}\label{fig:BS-BB}
\end{figure}

We start by reviewing the approximate static metric - which we already obtained in \cite{Haddad:2012ss} - that describes a test black string dangling from the boundary to the horizon of the planar Schwarzschild-AdS$_{5}$ black hole, see Fig.\ref{fig:BS-BB}. This is a solution of the Einstein equations with negative cosmological constant in $5$ dimensions,
\begin{equation}
R_{\mu\nu}-\frac{1}{2}R g_{\mu\nu}-\frac{6}{R^{2}_{\text{AdS}}}g_{\mu\nu}=0\,,
\end{equation}
where $R_{\text{AdS}}$ is the radius of curvature of AdS$_{5}$. The solution \cite{Haddad:2012ss} describes how to immerse a probe black string, with local metric,
\begin{eqnarray}
ds^2=dz^{2}-\left(1-\frac{r_{0}}{r}\right)dv^2+2dvdr+r^{2}d\Omega_{2}^{2}\,,
\end{eqnarray}
in the planar Schwarzschild-AdS$_{5}$ geometry. The planar geometry can be written more conveniently (in coordinates adapted to the black string) as
\begin{eqnarray}\nonumber\label{black brane}
ds^2&=&\left[\frac{R^{2}_{\text{AdS}}}{z^2f(z)}-\frac{z^2}{R^{2}_{\text{AdS}}}\left(\frac{r\partial_{z}f(z)}{2f(z)}\right)^{2}\right]dz^{2}-\frac{z^2}{R^{2}_{\text{AdS}}}\left(\frac{r\partial_{z}f(z)}{2f(z)}\right)\left[2\sqrt{f(z)}dvdz-2drdz\right]\\&+&\frac{z^2}{R^{2}_{\text{AdS}}}\left[-f(z)dv^2+2\sqrt{f(z)}dvdr+r^{2}d\Omega_{2}^{2}\right]\,,
\end{eqnarray}
where the red-shift function of the black brane is,
\begin{eqnarray}
f(z)=1-\frac{\mu}{z^4}\,.
\end{eqnarray}
By the coordinate change $v=t+r/\sqrt{f(z)}$ one can go back from (\ref{black brane}) to the familiar form of the black brane geometry,
\begin{eqnarray} 
ds^2=\frac{R^{2}_{\text{AdS}}dz^{2}}{z^2f(z)}+\frac{z^2}{R_{\text{AdS}}^{2}}\left[-f(z)dt^2+dr^2+r^{2}d\Omega_{2}^{2}\right]\,.
\end{eqnarray}
Before writing down the solution we want to emphasize that in our solution the black string horizon is much smaller than the radius of curvature of AdS$_{5}$, namely, $r_{0}/R_{\text{AdS}}<<1$, where the last plays the role of the small parameter in our solution, and which tells that our black string is a test object. We assume also that the planar black hole horizon is of the same order of magnitude as the AdS$_{5}$ radius, that is, $\mu^{1/4}\sim R_{\text{AdS}}$. Note that the last two assumptions imply that the temperature of the boundary black hole, $T_{\text{B.H}}\sim 1/r_{0}$, is much higher than the temperature of the surrounding plasma, $T_{\text{plasma}}\sim 1/R_{\text{AdS}}$. Namely, $T_{\text{B.H}}>>T_{\text{plasma}}$.

The solution \cite{Haddad:2012ss}, expanded up to second order in derivatives around an arbitrary $z=\text{constant}$ surface (we denote the surface by $z=z_c$ and assume that $z_c>\mu^{1/4}$) is given by the  following metric,\footnote{This metric is of the same structure as the metrics found in the Fluid-Gravity correspondence \cite{Bhattacharyya:2008jc} and in the Blackfolds approach \cite{Emparan:2009at,Emparan:2010ch} in the sense that it is derived by using a derivative expansion method.}
\begin{eqnarray}\nonumber\label{metric}
ds^2&=&\left[\frac{R^{2}_{\text{AdS}}}{z^2f(z)}-\frac{z^2}{R^{2}_{\text{AdS}}}\left(\frac{r\partial_{z}f(z)}{2f(z)}\right)^{2}\right]dz^{2}-\frac{z^2}{R^{2}_{\text{AdS}}}\left(\frac{r\partial_{z}f(z)}{2f(z)}\right)\left[2\sqrt{f(z)}dvdz-2drdz\right]\\\nonumber&+&\frac{z^2}{R^{2}_{\text{AdS}}}\left[-f(z)\left(1-\frac{r_{0}(z)}{r}\right)dv^2+2\sqrt{f(z)}dvdr+r^{2}d\Omega_{2}^{2}\right]
+\epsilon^{2} h^{(2)}_{\mu\nu}(r)dx^{\mu}dx^{\nu}+O(\epsilon^{3})\\
\,
\end{eqnarray} 
where $\epsilon$ is a formal parameter that counts the number of derivatives with respect to $z$, or equivalently counts the number of powers of $\frac{r_{0}}{R_\text{AdS}}$ in each term. In other words, as is explained in \cite{Haddad:2012ss}, each derivative with respect to $z$ brings out a factor of $\frac{r_{0}}{R_\text{AdS}}$. The non-vanishing components of $h^{(2)}_{\mu\nu}(r)$ are  
\begin{eqnarray}\nonumber\label{corrections}
&h^{(2)}_{vv}(r)&=
4\frac{\mu r_{0}^2}{R^{6}_{\text{AdS}}}\left[\left(1+\frac{\mu}{2z_{c}^{4}}\right)\frac{r}{r_{0}}+f(z_{c})\left(1-\frac{3r_{0}}{2r}\right)\log\frac{r}{r_{0}}\right]\,,\\
&h^{(2)}_{vr}(r)&=-2\frac{\mu r_{0}^2}{R^{6}_{\text{AdS}}}f(z_{c})^{-\frac{1}{2}}\left[\frac{\mu}{z_{c}^{4}}\frac{r}{r_{0}}+f(z_{c})\log\frac{r}{r_{0}}+\frac{1}{2}f(z_{c})\right]\,,\\\nonumber
&h^{(2)}_{zz}(r)&=8\frac{\mu}{z_{c}^{4}}\frac{r_{0}^2}{R^{2}_{\text{AdS}}}f(z_{c})^{-1}\left[\frac{r}{r_{0}}+\log\frac{r}{r_{0}}\right]\,,
\end{eqnarray} 
and the $z$-dependent radius of the black string is given by 
\begin{eqnarray}
r_{0}(z)=2M\sqrt{f(z)}\,.
\end{eqnarray}
Note that $2M$ is the radius of the boundary black hole since $r_{0}(z)\rightarrow 2M$ as $z\rightarrow \infty$. One should understand all quantities, which appear in the above solution, as expanded in derivatives (with respect to $z$) around the arbitrary surface $z=z_c$. For example, 
\begin{eqnarray}
f(z)=f(z_{c})+\epsilon \left(z-z_{c}\right)\partial_{z}f(z_{c})+\epsilon^{2} \frac{\left(z-z_{c}\right)^{2}}{2}\partial^{2}_{z}f(z_{c})+O(\epsilon^{3})\,.
\end{eqnarray} 
It is worth mentioning that this solutions is regular everywhere in the Lorentzian section, and in special, the black string caps off smoothly at the planar black hole horizon. However, the solution is conical in the Euclidean section at the point where the black string intersects the planar black hole, and that is because each object has a different temperature. Moreover, the above solution satisfies the correct boundary conditions, see \cite{Haddad:2012ss} for details. In particular, for large values of $z_{c}$ the solution reduces to the familiar AdS$_{5}$ black string, and hence it induces (up to a conformal factor) a $4-$dim Schwarzschild geometry on the AdS$_{5}$ boundary.

\section{Holographic Stress Tensor}
\label{sec:2}
In this section we take the above bulk solution and compute from it the holographic stress tensor (the boundary stress tensor) using the prescription of \cite{Balasubramanian:1999re}. That is, we use the well-known formula \footnote{Our notations introduce a minus sign difference from \cite{Balasubramanian:1999re} in $\Theta_{ab}$. We use the notations of \cite{Wald:1984}, not of \cite{Weinberg:1972}, and therefore we have also a minus sign difference in the Riemann tensor with respect to \cite{Balasubramanian:1999re,Weinberg:1972}.}
\begin{equation}\label{Nidal1 stress}
T_{ab}=\frac{R_\text{AdS}}{16\pi G_{5}}\left[E_{ab}-\frac{2}{R_\text{AdS}}\left(\Theta_{ab}-\Theta\gamma_{ab}\right)-\frac{6}{R_\text{AdS}^{2}}\gamma_{ab}\right]\,,
\end{equation}   
where $\gamma_{ab}$ is the induced metric on the $z=z_{c}$ surface, $\Theta_{ab}=(\nabla_{a}n_{b}+\nabla_{b}n_{a})/2$ is the extrinsic curvature of that surface - $n_{a}$ is an outward pointing normal vector to the surface - and $E_{ab}$ is the Einstein tensor with respect to $\gamma_{ab}$.
The key step in our calculation of the boundary stress tensor is to notice that if we multiply both sides of (\ref{Nidal1 stress}) by the quantity $r_{0}^{2}$ then it becomes manifest that the three terms on the right hand side are of different orders in the small parameter of our system, $r_{0}/R_\text{AdS}$ . Therefore, one can write down
\begin{equation}\label{Nidal2 stress}
T_{ab}=\frac{R_\text{AdS}}{16\pi G_{5}}\left[E_{ab}-\frac{2\epsilon}{R_\text{AdS}}\left(\Theta_{ab}-\Theta\gamma_{ab}\right)-\frac{6\epsilon^{2}}{R_\text{AdS}^{2}}\gamma_{ab}\right]\,,
\end{equation}
where the $\epsilon$ parameter, as discussed above, indicates of what order each term is (see Appendix \ref{app:details} for more details on the last step). Hence, upon pluging our solution (\ref{metric}) into eq.(\ref{Nidal2 stress}), and taking thereafter the limit $z_{c}\rightarrow\infty$ one obtains the following stress tensor: 
\begin{equation}\label{Nidal3 stress}
T_{ab}dx^adx^b=\frac{\epsilon^{2}\mu}{16\pi G_{5}R_\text{AdS}^{3}z_{c}^{2}}\left[a(r)dv^2+2b(r)dvdr+c(r)dr^2+d(r)r^{2}d\Omega^{2}_{2}\right]\,,
\end{equation}
with 
\begin{eqnarray}\nonumber\label{abcd}
&a(r)&=3-5\frac{2M}{r}+2\frac{\left(2M\right)^4}{r^4}
\,,\\
&b(r)&=-3+2\frac{2M}{r}+2\frac{\left(2M\right)^2}{r^2}+2\frac{\left(2M\right)^3}{r^3}\,,\\\nonumber
&c(r)&=4\left(1-\frac{\left(2M\right)^2}{r^2}\right)\,,\\
&d(r)&=1-4\frac{\left(2M\right)^3}{r^3}\,.\nonumber
\end{eqnarray}
Note that, as it should, the stress tensor is proportional to $z_{c}^{-2}$, gauranteeing a finite mass density on the boundary. Since that is the end of our calculation we will put now the formal parameter $\epsilon$ back to unity. We also will take the boundary metric to be the Schwarzschild metric, and so we need to make the following conformal transformation $\gamma_{ab}\rightarrow \frac{R_\text{AdS}^{2}}{z_{c}^2}\gamma_{ab}$. As a result, the stress tensor will transform as $T_{ab}\rightarrow \frac{z_{c}^2}{R_\text{AdS}^{2}}T_{ab}$, and so it will take the following desired form,
\begin{equation}\label{Nidal4 stress}
T_{ab}dx^adx^b=\frac{\mu}{16\pi G_{5}R_\text{AdS}^{5}}\left[a(r)dv^2+2b(r)dvdr+c(r)dr^2+d(r)r^{2}d\Omega^{2}_{2}\right]\,,
\end{equation}
which is the main result of our work.
 
\section{Properties of the Stress Tensor}
\label{sec:3} 
The stress tensor found above should be identified according to the AdS/CFT correspondence with the expectation value of the stress tensor, in the Unruh vacuum,\footnote{This should be identified with the Unruh vacuum because the boundary black hole has a temperature which is higher than that of the plasma and so one expects a situation where there is only outward flux of positive energy. Typically, the Unruh state which is more common in the literature is the one where there is initially empty space outside the collapsing body. What we have at hand, however, is a slightly different Unruh state. It is the steady state for a collapsing black hole which was initially surrounded by thermal plasma.} of $\mathcal{N}=4$, $SU(N)$ super Yang-Mills theory on the $4-$dim Schwarzschild background, at large $N$ and large 't Hooft coupling (see \cite{Hubeny:2009ru}),
\begin{equation}\label{}
\left\langle T_{ab}\right\rangle_{\text{ren}}=T_{ab}\,.
\end{equation} 
The result for the stress tensor obtained above is at leading order in the large $N$ limit. By looking at dimensionful factor that multiplies the stress tensor (\ref{Nidal4 stress}), and by using the relation $\frac{R_{\text{AdS}}^{3}}{G_5}=\frac{2N^2}{\pi}$ one can rewrite the stress tensor as,
\begin{eqnarray}\nonumber\label{}
T_{ab}dx^adx^b&=&\frac{\mu N^2}{8\pi^{2} R_\text{AdS}^{8}}\left[a(r)dv^2+2b(r)dvdr+c(r)dr^2+d(r)r^{2}d\Omega^{2}_{2}\right]
\,\\
&=&O(N^2)\,.\\\nonumber
\end{eqnarray}
As expected from a classical bulk solution, it gives the leading order part, $O(N^2)$, of the CFT stress tensor \cite{Hubeny:2009ru,Figueras:2011va}. Eventhough this stress tensor is non-vanishing at this leading order in $N$, we will see later that it contains no flux terms, or in other words, there is no transport of energy from the black hole to infinity at this order \cite{Emparan:2002px,Tanaka:2002rb,Fitzpatrick:2006cd,Gregory:2008br}. As explained for example in \cite{Fischetti:2012ps}, for black droplets heat transport occurs as a result of a bulk quantum process (Hawking radiation in the bulk) which appears at order $O(N^0)$.

\subsubsection*{Finite and covariantly conserved}
\label{sec:finite}
The Eddington-Finklstein coordinates, $(\upsilon,r)$, used above to write the boundary metric and the stress tensor, make regularity and finiteness of tensors on the future horizon manifest. Note that the stress tensor (\ref{Nidal4 stress}) is finite everywhere. In particular, it is finite at long distances from the boundary black hole, at $r>>2M$, and at the future horizon of the boundary black hole, that is, at $r=2M$.  The $(\upsilon,r)$ coordinates are, nevertheless, not appropriate for treating the past horizon. To talk about the past horizon one can, for example, use the $(u,r)$ coordinates instead, where as usual, $u=t-r_{*}$ and the tortoise coordinate is $r_*=r+2M\log[r/2M-1]$. One can easily check that the stress tensor remains finite in the $(u,r)$ coordinates, implying finiteness at the past horizon as well. In short, we have found that at this leading order in $N$ the expectation value  of the stress tensor in the Unruh state is finite everywhere at and outside the horizon.

In this regard, we would like to comment that eventhough the Unruh state is commonly known (in a free field theory) to be divergent on the past horizon, we find here that (at strong coupling and large $N$) it is finite there at leading order in $N$. Similar results for the Unruh and Boulware states at large $N$ and strong coupling were found in \cite{Figueras:2011va,Figueras:2013jja,Fischetti:2013hja}. In this concern, the authors in \cite{Figueras:2011va} expect that the stress tensor will regain the divergency on the past horizon when subleading terms in $N$ are included.

As for covariant conservation, one can check, by direct calculation, that our boundary stress tensor is covariantly conserved with respect to the boundary metric (the Schwarzschild metric), 
\begin{equation}\label{}
\nabla^{a}T_{ab}=0\,.
\end{equation}
This is, in fact, in accord with what one expects from the dual quantum field theory point of view; the renormalized stress tensor $\left\langle T_{ab}\right\rangle_{\text{ren}}$ is covariantly conserved, since it is derived from an effective action \cite{Birrell:1982a}. In \cite{Christensen:1977a} the authors start their search for the general form of the renormalized stress tensor (in the Schwarzschild background) by looking for a general solution for the above conservation equation, under the assumption of staticity and spherical symmetry. In our case, however, it is to be noted that we derive the conservation equation instead of assuming it.

\subsubsection*{The stress tensor at infinity}
Far away from the boundary black hole, that is, for $r>>2M$, the stress tensor (\ref{Nidal4 stress}) reduces to the stress tensor of thermal plasma at temperature $T=\frac{\mu^{1/4}}{\pi R_{\text{AdS}}^{2}}$, 
\begin{equation}\label{}
T_{ab}dx^adx^b=\frac{\mu}{16\pi G_{5}R_\text{AdS}^{5}}\left[3dv^2-6dvdr+4dr^2+r^{2}d\Omega^{2}_{2}+O\left(M/r\right)\right]\,.
\end{equation} 
If we perform a boundary coordinate transformation to the Schwarzschild coordinates, given by $v=t+r_*=t+r+2M\log[r/2M-1]$, then the above asymptotic stress tensor will take the more familiar form,
\begin{equation}\label{}
T_{ab}dx^adx^b=\frac{\mu}{16\pi G_{5}R_\text{AdS}^{5}}\left[3dt^2+dr^2+r^{2}d\Omega^{2}_{2}+O\left(M/r\right)\right]\,.
\end{equation} 
However, as the parameter $M$ which characterizes the boundary black hole does not appear in the asymptotic expression for the stress tensor we must conclude that the thermal plasma at infinity (dual to the planar black hole in the bulk) is not influenced by the black hole at this leading order $O(N^2)$. Nevertheless, as the fall-off of the next-to-leading order components of the stress tensor goes like $M/r$, one concludes that the interaction between the black hole and the plasma is not as weak as is typical for the black droplet phase. In \cite{Figueras:2011va} it was found that the fall-off is $1/r^5$, which clearly implied weak interaction. The reason for this difference is that our black droplet is a critical one, as it touches the planar black hole in the bulk, whereas the droplets studied in \cite{Figueras:2011va} are strictly above the Poincare horizon. This means that our black droplet interacts stronger with planar black hole than the droplets of \cite{Figueras:2011va}, and this is reflected in the boundary theory by having a stronger interaction between the plasma at infinity and the boundary black hole in our case.

\subsubsection*{Trace anomaly}
Renormalized stress tensors for weakly coupled conformal field theories have been extensively studied (see \cite{Birrell:1982a}), and they are known to have a trace anomaly. According to \cite{Balasubramanian:1999re,Henningson:1998gx} the trace anomaly in the strongly coupled case takes exactly the same form as in the weakly coupled case. In $4$ spacetime dimensions, which is the case of our interest, it takes the following form, 
\begin{equation}\label{anomaly}
T^{a}_{a}=-\frac{R_\text{AdS}^{3}}{8\pi G_{5}}\left[-\frac{1}{8}R^{ab}R_{ab}+\frac{1}{24}R^{2}\right]\,.
\end{equation} 
For our system, we have found that the trace anomaly does not appear up to the order we are working to. That is, our stress tensor (\ref{Nidal4 stress}) gives,
\begin{equation}\label{trace}
T^{a}_{a}=O(\epsilon^{3})=O(M^3/R_\text{AdS}^3)\,.
\end{equation} 
Our result (\ref{trace}) is consistent with eq.(\ref{anomaly}) because up to the order we are working to (we did not compute the back-reacted metric) the boundary metric is pure Schwarzschild ($R_{ab}=0$) and therefore eq.(\ref{anomaly}) will give $T^{a}_{a}=0$, confirming our result. Non-zero contributions to eq.(\ref{anomaly}) may appear only after computing the back-reacted metric. The corrections to the Schwarzschild metric due to the backreaction will be of order $M^2/R_\text{AdS}^2$, and so, eq.(\ref{anomaly}) will give $T^{a}_{a}=O(M^4/R_\text{AdS}^4)$, which means that non-zero contributions to the trace-anomaly, if any, will be of order $O(M^4/R_\text{AdS}^4)$ at least.

\subsubsection*{Negative energy density near the horizon}
Here we are going to show that the energy density of our system displays a crucial ingredient of stress tensors that describe Hawking radiation. Namely, we are going to show that near the horizon of the black hole there is a region with a negative energy density while near infinity the energy density is positive. This is a manifestation of particle creation. A pair of particles is created near the black hole horizon, one particle with negative energy enters the black hole horizon while its partner with positive energy escapes to infinity. In particular, that is the way a black hole is expected to lose its mass and evaporate; by absorbing negative energies.

\begin{figure}[ht]
\begin{center}{}
\includegraphics[width=0.6\textwidth]{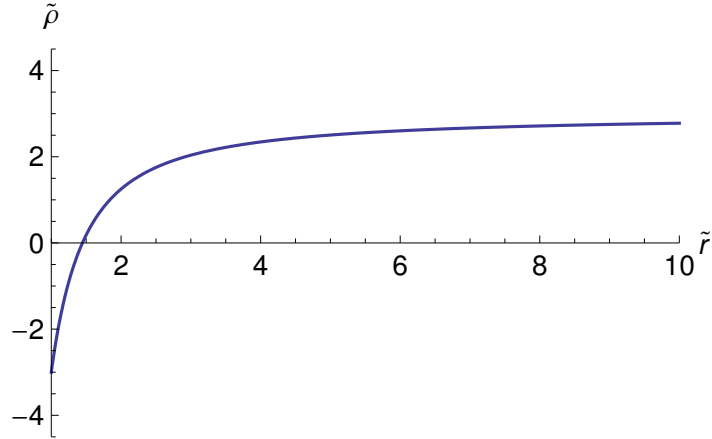}
\end{center}
\caption{A plot of the normalized energy density, defined by $\tilde{\rho}=\frac{16\pi G_{5}R_\text{AdS}^{5}}{\mu}\rho$, with respect to $\tilde{r}=r/2M$. Note that the energy density is finite at both the horizon and at infinity. Note furthermore the negative energy denisty in the region $\tilde{r}\in [1,\approx 1.5)$.
}\label{fig:densitystrong} 
\end{figure}

Let us look at the energy density of the boundary theory, 
\begin{equation}\label{}
\rho=T_{ab}u^a u^b\,,
\end{equation}  
where we will take, $u^a=(1-2M/r)^{-1/2}\delta^{a}_\upsilon$, the $4-$velocity of a static observer. This gives us,
\begin{equation}\label{}
\rho=\frac{\mu}{16\pi G_{5}R_\text{AdS}^{5}}\left[3-2\frac{2M}{r}-2\frac{(2M)^2}{r^2}-2\frac{(2M)^3}{r^3}\right]\,.
\end{equation} 
In Fig.\ref{fig:densitystrong}  we have plotted the energy density, and there one can see that in the region near the black hole horizon, $r\in[2M,r_{1}\approx 3M)$, the energy density is negative, while in the region $r\in(r_{1}\approx 3M,\infty)$ the energy density is positive.

\subsubsection*{No flux at leading order in $N$}

Here we compute the energy-momentum current $J_{a}$, defined as (see\cite{Wald:1984})
\begin{equation}
J_{a}=-T_{ab}u^b\,,
\end{equation} 
where $u^{a}$ is the $4-$velocity of an observer in the boundary spacetime. Again, let us take $u^a=(1-2M/r)^{-1/2}\delta^{a}_\upsilon$. It is straightforward to check that the stress tensor (\ref{Nidal4 stress}) gives,
\beq
J^a=\frac{\rho}{(1-2M/r)^{1/2}}\delta^{a}_\upsilon\,.
\eeq
In special, this gives that $J^r=0$, which tells that there is no radial flux. Note that this is equivalent to showing that $T^{r}_{\upsilon}=0$. As said before this result agrees with recent results for similar calculations at strong coupling, see \cite{Figueras:2011va,Figueras:2013jja,Fischetti:2013hja}. In reference \cite{Fischetti:2013hja} the authors gave the black droplet phase the name "jammed phase", since it behaves more like a solid, with no flow, rather than a fluid (black funnels). Yet, one expects that flux terms will appear in subleading orders in $N$, and they will lead to an exchange of heat between the boundary black hole and the surrounding plasma.

\section{Final Comments}
\label{sec:comments}
Most works so far on constructing black droplet solutions - and more generally on constructing static black hole solutions in AdS that induce black hole metrics on the boundary - relyed on numerical calculations (see \cite{Hubeny:2009ru} and references therein). In the work \cite{Haddad:2012ss} we provided the first analytical example of a black droplet in $AdS_{5}$ spacetime and here, in this work, we provide its dual stress tensor. The construction of the analytical solution of the black droplet is due to the small parameter $r_{0}/R_\text{AdS}$ of our sysytem which allowed for the perturbative construction. The calculation of the holographic stress tensor was straightforward and we obtained a simple and compact stress tensor which has the following important features: The stress tensor is static, covariantly conserved, regular on both future and past horizons, it gives a negative energy density near the black hole horizon and a positive energy density at infinity, and it contains no energy transfer terms at this (leading) order. These features agree with expectations and results on similar settings \cite{Hubeny:2009ru,Fischetti:2012ps,Figueras:2011va}. It is worth to conclude by the comment that this agreement gives further evidence that the analytical solution found in \cite{Haddad:2012ss} indeed describes a black droplet. 
    
\appendix
\section{Details on the calculation of the stress tensor}\label{app:details}   
In this appendix I am going to explain the passage from eq.(\ref{Nidal1 stress}) to eq.(\ref{Nidal2 stress}). Remember first that $\epsilon$ is a formal parameter which is inserted wherever there is a factor $\frac{r_{0}}{R_{\text{AdS}}}$; it simply counts the number of powers of $\frac{r_{0}}{R_{\text{AdS}}}$ and helps in organizing the calculations. 

Upon multiplying eq.(\ref{Nidal1 stress}) by $r_{0}^2$ one obtains,
\begin{equation}\label{1}
r_{0}^{2}T_{ab}=\frac{R_\text{AdS}}{16\pi G_{5}}\left[r_{0}^{2}E_{ab}-2\frac{r_{0}}{R_\text{AdS}}r_{0}\left(\Theta_{ab}-\Theta\gamma_{ab}\right)-6\frac{r_{0}^{2}}{R_\text{AdS}^{2}}\gamma_{ab}\right]\,.
\end{equation}
Now, as said above, wherever we see a factor $\frac{r_{0}}{R_{\text{AdS}}}$ we insert $\epsilon$, and so we obtain,
\begin{equation}\label{2}
r_{0}^{2}T_{ab}=\frac{R_\text{AdS}}{16\pi G_{5}}\left[r_{0}^{2}E_{ab}-2\epsilon\frac{r_{0}}{R_\text{AdS}}r_{0}\left(\Theta_{ab}-\Theta\gamma_{ab}\right)-6\epsilon^{2}\frac{r_{0}^{2}}{R_\text{AdS}^{2}}\gamma_{ab}\right]\,.
\end{equation}
To show that every thing is consistent let us see now how the calculation proceed. Look at the $3$ dimensionless quantities $r_{0}^{2}E_{ab}$, $r_{0}\left(\Theta_{ab}-\Theta\gamma_{ab}\right)$, and $\gamma_{ab}$ in the above expression and expand them in derivatives as well (as is done to all quantities in such calculations),
\begin{eqnarray}\nonumber\label{3}
&r_{0}^{2}E_{ab}&=e^{(0)}_{ab}(x)+\epsilon \left(z-z_{c}\right)e^{(1)}_{ab}(x)+\epsilon^{2} \frac{\left(z-z_{c}\right)^{2}}{2}e^{(2)}_{ab}(x)+O(\epsilon^{3})\,,\\
&r_{0}\left(\Theta_{ab}-\Theta\gamma_{ab}\right)&=\theta_{ab}^{(0)}(x)+\epsilon \left(z-z_{c}\right)\theta_{ab}^{(1)}(x)+\epsilon^{2} \frac{\left(z-z_{c}\right)^{2}}{2}\theta_{ab}^{(2)}(x)+O(\epsilon^{3})\,,\\\nonumber
&\gamma_{ab}&=\gamma_{ab}^{(0)}(x)+\epsilon \left(z-z_{c}\right)\gamma_{ab}^{(1)}(x)+\epsilon^{2} \frac{\left(z-z_{c}\right)^{2}}{2}\gamma_{ab}^{(2)}(x)+O(\epsilon^{3})\,,
\end{eqnarray} 
where $x$ denotes the boundary coordinates, and where, for example, $e^{(0)}_{ab}(x)=\left[r_{0}^{2}E_{ab}\right]_{z=z_c}$, $e^{(1)}_{ab}(x)=\left[\partial_{z}\left(r_{0}^{2}E_{ab}\right)\right]_{z=z_c}$, and $e^{(2)}_{ab}(x)=\left[\partial_{z}^{2}\left(r_{0}^{2}E_{ab}\right)\right]_{z=z_c}$. This shows the way the calculation is organized, and it makes it clear that the addition of the $\epsilon$'s in going from eq.(\ref{Nidal1 stress}) to eq.(\ref{Nidal2 stress}) is indeed consistent.

\end{document}